\documentclass[epj]{svjour}

\usepackage{graphicx}
\usepackage{amsmath}
\usepackage{amsbsy}

\newcommand{\kk}{{\bf k}}

\newcommand{\xx}{{\bf x}}
\newcommand{\yy}{{\bf y}}

\newcommand{\BE}{\begin{equation}}
\newcommand{\EE}{\end{equation}}
\newcommand{\BA}{\begin{eqnarray}}
\newcommand{\EA}{\end{eqnarray}}
\newcommand{\Tr}{\mathop{\mathrm{Tr}}}
\newcommand{\kGL}{\kappa_{\mathrm{GL}}}

\begin{document}

\title{Gaussian Effective Potential and superconductivity}

\author{M. Camarda\inst{1,2} 
\and G. G. N. Angilella\inst{1,2} 
\and R. Pucci\inst{1,2} 
\and F. Siringo \inst{1,3}}

\institute{Dipartimento di Fisica e Astronomia, Universit\`a di
   Catania, Via S. Sofia, 64, I-95123 Catania, Italy
\and
Istituto Nazionale per la Fisica della Materia, UdR Catania
\and
Istituto Nazionale di Fisica Nucleare, Sezione di Catania
}

\date{\today}

\abstract{%
The Gaussian Effective Potential in a fixed transverse unitarity
gauge is studied for the static 
three-dimensional U(1) scalar electrodynamics 
(Ginzburg-Landau phenomenological theory of superconductivity). 
In the broken-symmetry phase
the mass of the electromagnetic field (inverse penetration depth) and the
mass of the scalar field (inverse correlation length) are both determined
by solution of the coupled variational equations. At variance with
previous calculations, the choice of a fixed unitarity gauge prevents
from the occurrence of any unphysical degree of freedom. The theory
provides a nice interpolation of the experimental data when approaching the
critical region, where the standard mean-field method is doomed to
   failure.
}

\maketitle

\section{Introduction}

Since the discovery of high $T_c$ superconductors several
unconventional models have been proposed in order to describe
the unusual properties of cuprates. 
In particular, the strong electron-electron coupling characterizing
   such materials requires new theoretical methods beyond the standard mean 
field approach. On the other hand the general phenomenology is
still well described by the standard Anderson-Higgs mechanism:
the supercurrent is carried by pairs of charged fermions whose
non-vanishing expectation value breaks the gauge symmetry, thus
   endowing the gauge bosons with a mass. Thus the standard 
Ginzburg-Landau (GL) effective lagrangian still provides the best 
framework for a general description of the high-$T_c$ cuprates phenomenology.
Moreover, as the GL action can be seen as a power expansion of
the exact action around the critical point, the GL action must be
recovered by any microscopic theory at least around the transition.
Thus, regardless of the nature of the pairing mechanism,
the GL action is a sound starting point for a general description
of the high-$T_c$ materials. Of course we cannot trust the mean-field
approach to the GL effective theory, and we expect that many
unconventional properties are connected with the breaking down of
the simple mean field picture. 
Quite recently, the GL model has been extensively studied both
   theoretically \cite{Herbut:96,Tesanovic:99} and numerically
   \cite{Nguyen:99} in order to clarify the universality class of the
   superconducting transition and the role of the critical
   fluctuations in the high-$T_c$ cuprates, as well as the order of
   the transition itself \cite{Mo:02}.

Actually the high-$T_c$ cuprate superconductors
are characterized by a very small correlation length $\xi$ which
allows the experimentalists to get closer to the critical point
where the thermal fluctuations cannot be neglected and the mean
field approximation is doomed to fail \cite{Larkin:02}.
As far as we know there is no full evidence that the universal critical 
behaviour has been reached in any real sample \cite{Schneider:00}, but it
is out of doubt that an intermediate range of temperature is
now accessible, where thermal fluctuations are not negligible
even if the sample is still out of the truly critical regime.
Thus, in order to describe some unconventional properties of
the high-$T_c$ superconductors, we need to incorporate the role
of thermal fluctuations, but unfortunately we cannot rely
on the standard renormalization group methods which would only
describe the universal limiting behaviour that could not be observed
yet in any real sample. We need some kind of interpolation scheme
for the non-universal regime where the behaviour depends on the
physical parameters of the sample, and we would prefer a 
non-perturbative method in order to deal with any strong coupling.

In this paper we study the Gaussian fluctuations by means of a variational
method, the Gaussian Effective Potential (GEP), which has been discussed
by several authors as a tool for describing the breaking of symmetry
in a simple scalar theory \cite{Stevenson:85,Cuccoli:95}.
As a toy model for electro-weak interactions,
the scalar electrodynamics in four dimensions has been studied by
Iba\~nez-Meier \emph{et al.}~\cite{IbanezMeier:96} who computed the
   GEP by use of a general 
covariant gauge. However their method gives rise to an unphysical,
and undesirable, degree of freedom.

We compute the GEP for the U(1) scalar electrodynamics
in three space dimensions where it represents the standard static
GL effective model of superconductivity. In order to make evident
the physical content of the theory, thus avoiding the presence of
unphysical degrees of freedom, we work in unitarity gauge. This
has been shown to be formally equivalent to a full
gauge-invariant method once all the gauge degrees of freedom have
been integrated out \cite{Mansfield:86}. The variational method
provides a way to evaluate both the correlation length $\xi$ and the
penetration depth $\ell$ as a solution of coupled equations.
The GL parameter $\kGL=\ell/\xi$ is found to be strongly temperature
dependent in contrast to the simple mean-field description
   \cite{Helfand:66}. On the 
other hand the model predictions are in perfect agreement with some 
recent experimental data \cite{Brandstatter:94},
which can be nicely interpolated by our variational calculation.

The comparison with the experimental data is of special importance
as it provides a test for the GEP variational method itself.
The predictions of the method in 3+1 dimensions, in the
context of electro-weak interactions, have been discussed by several
authors \cite{Stevenson:85,IbanezMeier:96,Stancu:90,Consoli:00},
but no real comparison with experimental data will be achievable
until the detection of the Higgs boson. Thus high-$T_c$ cuprate
superconductors represent our next best choice in order to test the
reliability of the method.

The paper is organized as follows: in section~\ref{sec:GL} the GL action and
partition function are introduced and discussed within the unitarity gauge;
in section~\ref{sec:gaussian} the Gaussian variational method is applied 
to the three-dimensional GL model, and the coupled variational
equations are derived; finally, in section~\ref{sec:comparison} the
   phenomenological 
predictions of the method are compared with some experimental data
for the GL parameter.

\section{The GL action in unitarity gauge}
\label{sec:GL}

Let us consider the standard static GL action \cite{Kleinert:89}
\BA
S &=& \int d^{3}x \left[\frac{1}{4} F_{\mu \nu}F^{\mu
\nu}+\frac{1}{2}(D_{\mu}\phi)^{*}(D^{\mu}\phi) \right.\nonumber\\
&&+
\left.\frac{1}{2}m^{2}_{B}\phi^{*}\phi+
\lambda_{B}(\phi^{*}\phi)^{2}\right].
\label{gl}
\EA
Here $\phi$ is a complex (charged) scalar field, whose covariant
derivative is defined as
\BE
D_{\mu}\phi = \partial_{\mu}+ie_{B} A_{\mu}
\label{derivative}
\EE
and $\mu,\nu=1,2,3$ run over the three space dimensions.
The magnetic field components 
$F_{\mu\nu}=\partial_\mu A_\nu-\partial_\nu A_\mu$ satisfy
\BE
\frac{1}{2}F_{\mu \nu}F^{\mu \nu} =\vert{\nabla}\times
{\mathbf A}\vert^2
\EE
and the partition function is defined by the functional integral
\BE
Z=\int D[\phi,\phi^*,A_\mu]e^{-S}.
\label{z}
\EE
We may assume a transverse gauge $\nabla\cdot{\bf A}=0$, and then
switch to unitarity gauge in order to make $\phi$ real.
Let us define two real fields $\rho$ and $\gamma$ according to
$\phi=\rho e^{i\gamma}$.
The unitarity gauge is recovered by the gauge transformation
\BE
{\bf A} \to
{\bf A}-\frac{1}{e_{B}}\nabla \gamma(x)
\EE
and the original transverse vector field ${\bf A}_\perp$ acquires
a longitudinal component ${\bf A}_L$ proportional to $\nabla\gamma$.
Thus the original measure in Eq.~(\ref{z}) becomes
\BA
\int D[\phi,\phi^*, {\bf A}_\perp] &=&
\int D[\gamma]\rho D[\rho] D[{\bf A}_\perp] \nonumber\\
&\to&
\mathrm{const}\times\int \rho D[\rho] D[{\bf A}_L] D[{\bf A}_\perp]
\label{measure}
\EA
In unitarity gauge the action, Eq.~(\ref{gl}), now reads
\BA
S &=& \int d^{3} x \left\{
\frac{1}{2}(\nabla
\rho)^{2}+\frac{1}{2} m_{B}^{2}\rho^{2}
+\lambda_{B}\rho^{4} \right.\nonumber\\
&&\left. +\frac{1}{2} e_{B}^{2}\rho^{2} (A_L^{2}+A_\perp^2)+
\frac{1}{2}({\nabla} \times {\bf A}_\perp)^{2} \right\}
\label{su}
\EA
and the longitudinal field ${\bf A}_L$ may be integrated out exactly
yielding a constant factor and an extra $1/\rho$ factor 
for the measure (\ref{measure}). Finally, dropping the constant
factors, the partition function may be written as
\BA
Z &=& \int D[\rho, {\bf A}_\perp] 
\exp\left\{-\int d^{3} x \left[
\frac{1}{2}(\nabla\rho)^{2}+\frac{1}{2} m_{B}^{2} \rho^{2}
   \right.\right.\nonumber\\
&&+\left.\left.\lambda_{B}\rho^{4}+\frac{1}{2} e_{B}^{2}\rho^{2} A_\perp^2 +
\frac{1}{2}({\nabla} \times {\bf A}_\perp)^{2} \right]
\right\} .
\EA
We may enforce the transversal condition on the vector field
by a gauge fixing term in the action and, restoring
$\rho=\phi$, the action reads
\BA
S&=&\int d^{3} x \left[
\frac{1}{2}(\nabla\phi)^{2}+\frac{1}{2} m_{B}^{2} \phi^{2}
+\lambda_{B}\phi^{4}+\frac{1}{2} e_{B}^{2}\phi^{2} A^2 \right.\nonumber\\
&&+\left.
\frac{1}{2}({\nabla} \times {\bf A})^{2} 
+\frac{1}{2\epsilon} (\nabla\cdot{\bf A})^{2}
\right].
\label{su2}
\EA
The partition function is now expressed as a functional integral
over the real scalar field $\phi$ and the generic three-dimensional
vector field ${\bf A}$, with the extra prescription that the
parameter $\epsilon$ is set to zero at the end of the
calculation. Inserting a source term we may write
\BE
Z[j]=\int D[\phi,A_\mu] \exp\left\{-S+\int d^3 x j\phi\right\}
\label{z2}
\EE
with $S$ given by Eq.~(\ref{su2}).
The free energy (effective potential) follows by the Legendre
transformation
\BE
{\cal{F}}[\varphi]=-\ln Z+\int d^3 x j\varphi
\label{free}
\EE
where $\varphi$ is the average value of $\phi$
in presence of the source $j$. 
The superconducting phase is characterized by an
absolute minimum of $\cal{F}$ for $\varphi\not=0$.

\section{The Gaussian method}
\label{sec:gaussian}

From the partition function (\ref{z2}), the GEP may be evaluated
by the $\delta$ expansion method discussed in
   Refs.~\cite{IbanezMeier:96,Stancu:90}. In our case the GEP
   represents a variational 
estimate of the free energy (\ref{free}).

First we introduce a shifted field
\BE
\tilde{\phi}=\phi-\varphi
\EE
then we split the lagrangian into two parts
\BE
{\cal L}={\cal L}_0+{\cal L}_{int}
\EE
where ${\cal L}_0$ is the sum of two free-field terms
describing a vector field $A_\mu$ with mass $\Delta$ and
a real scalar field $\tilde \phi$ with mass $\Omega$:
\BA
{\cal L}_{0} &=& \left[
+\frac{1}{2}({\nabla} \times {\bf A})^{2} 
+\frac{1}{2} \Delta^{2} A_\mu A^\mu
+\frac{1}{2\epsilon}({\nabla}\cdot{\bf A})^{2} 
\right] \nonumber\\
&&+\left[
\frac{1}{2}({\nabla}\tilde{\phi})^{2}
+\frac{1}{2}\Omega^{2}\tilde{\phi}^{2}
\right] .
\EA
The interaction then reads
\BA
{\cal L}_{int}&=&
v_{0}+v_{1}\tilde{\phi}+v_{2}\tilde{\phi}^{2}+v_{3}\tilde{\phi}^{3}
+v_{4}\tilde{\phi}^{4} \nonumber\\
&&+\frac{1}{2}\left(e_{B}^{2}\varphi^{2}-\Delta^{2}\right)A_\mu A^\mu
\nonumber\\
&&+e_{B}^{2}\varphi A_\mu A^\mu\tilde{\phi}
+\frac{1}{2}e^{2}_{B}A_\mu A^\mu\tilde{\phi}^{2}
\EA
where
\begin{subequations}
\BA
v_{0} &=& \frac{1}{2}m_{B}^{2}\varphi^{2}+\lambda_{B}\varphi^{4}\\
v_{1} &=& m_{B}^{2}\varphi+4\lambda_{B}\varphi^{3}\\
v_{2} &=& \frac{1}{2}m_{B}^{2}+6\lambda_{B}\varphi^{2}-\frac{1}{2}\Omega^{2}\\
v_{3} &=& 4\lambda_{B}\varphi\\
v_{4} &=& \lambda_{B} .
\EA
\end{subequations}

We now expand the free energy to first order in ${\cal L}_{int}$ following
standard perturbation theory procedures. We obtain
\BA
{\cal F}[\varphi]&=&
\frac{1}{2}\Tr \ln\left[ g^{-1}(x,y)\right]
+\frac{1}{2}\Tr \ln\left[ G_{\mu\nu}^{-1}(x,y)\right] \nonumber\\
&+&\int d^3 x\left\{
v_0+v_2 g(x,x)+3v_4 g^2(x,x) \right.\nonumber\\
&&+\left.\frac{1}{2} e_B^2
\left(g(x,x)+\varphi^2-\Delta^2\right)G_{\mu\mu}(x,x)\right\}
\EA
where $g(x,y)$ is the free-particle propagator for the scalar field,
and $G_{\mu\nu}(x,y)$ is the free-particle propagator for the
vector field 
\BA
G_{\mu\nu}^{-1}(x,y) &=& \int \frac{d^3 k}{(2\pi)^3}
e^{-i\kk\cdot(\xx-\yy)}\left[\delta_{\mu\nu}(k^2+\Delta^2)
\right.\nonumber\\
&&+ \left.
\left(\frac{1}{\epsilon}-1\right) k_\mu k_\nu\right].
\EA
In the limit $\epsilon\to 0$, up to an additive constant
\BE
\Tr \ln\left[G_{\mu\nu}^{-1}(x,y)\right]=2{\cal V}
\int \frac{d^3 k}{(2\pi)^3} \ln (k^2+\Delta^2)
\EE
where $\cal V$ is the total volume.
Dropping all constant terms, the free energy density
$V_{\mathrm{eff}}={\cal F}/{\cal V}$ (effective potential) reads
\BA
V_{\mathrm{eff}}[\varphi] &=& I_{1}(\Omega)+2I_{1}(\Delta)+\nonumber\\
&&+ \left[
\lambda_{B}\varphi^{4}+\frac{1}{2}m_{B}^{2}\varphi^{2}+\frac{1}{2}
\left\{
m^{2}_{B}-\Omega^{2} \right.\right.\nonumber\\
&&+ \left.\left. 12\lambda_{B}\varphi^{2}+6\lambda_{B}I_{0}(\Omega)
\right\} I_{0}(\Omega) \right] \nonumber\\
&&+\left( e_B^{2}\varphi^{2}+e_B^{2}I_{0}(\Omega)-\Delta^{2} \right)
I_{0}(\Delta)
\label{veff}
\EA
where the divergent integrals $I_n$ are defined according to
\BE
I_0 (M)=\int\frac{d^3 k}{(2\pi)^3} \frac{1}{M^2+k^2}
\EE
\BE
I_1 (M)=\frac{1}{2}\int\frac{d^3 k}{(2\pi)^3} \ln(M^2+k^2)
\EE
and are regularized by insertion of a finite cut-off $\Lambda$.

The free energy (\ref{veff}) now depends on the mass parameters
$\Omega$ and $\Delta$. Since none of them was present in the original
GL action of Eq.~(\ref{su2}), the free energy should not depend on them,
and the {\it minimum sensitivity} method \cite{Stevenson:81} 
can be adopted in order to fix the masses: the free energy is required
to be stationary for variations of $\Omega$ and $\Delta$. On the other
hand the stationary point can be shown to be a minimum for the free
energy and the method is equivalent 
to a pure variational method \cite{IbanezMeier:96}.
At the stationary point the masses give the inverse correlation lengths
for the fields, the so called coherence length $\xi=1/\Omega$ and 
penetration depth $\ell=1/\Delta$.

The stationary conditions
\begin{subequations}
\BA
\frac{\partial V_{\mathrm{eff}}}{\partial\Omega^{2}} &=& 0\\
\frac{\partial V_{\mathrm{eff}}}{\partial\Delta^{2}} &=& 0
\EA
\end{subequations}
give two coupled gap equations:
\begin{subequations}
\label{gap12}
\BA
\label{gap1}
{\Omega}^{2} &=& 12 \lambda_{B}
I_{0}({\Omega})+m^{2}_{B}+12 \lambda_{B}\varphi^{2}+2
e^{2}_{B}I_{0}({\Delta}) \\
\label{gap2}
{\Delta}^{2} &= & e^{2}_{B}\varphi^{2}+e^{2}_{B}I_{0}({\Omega})
\EA
\end{subequations}
For any $\varphi$ value, Eqs.~(\ref{gap12}) must be solved 
numerically, and the minimum point values $\Omega$
and $\Delta$ must be inserted back into Eq.~(\ref{veff})
in order to get the gaussian free energy 
$V_{\mathrm{eff}}(\varphi)$
as a function of the order parameter 
$\varphi$.
For a negative and
small enough $m_B^2$, we find that $V_{\mathrm{eff}}$ has a minimum at a 
non zero value of $\varphi=\varphi_{\mathrm{min}}>0$, thus indicating that the
system is in the broken-symmetry superconducting phase.
Of course the masses $\Omega$, $\Delta$ only take their physical
value at the minimum of the free energy $\varphi_{\mathrm{min}}$.
This point may be found by requiring that
\BE
\frac{\partial V_{\mathrm{eff}}}{\partial\varphi^{2}}=0
\label{phimin}
\EE
where as usual the partial derivative is allowed as far as the
gap equations (\ref{gap12}) are satisfied \cite{Stevenson:85}.
The condition (\ref{phimin}) combined with the gap equation (\ref{gap1})
yields the very simple result
\BE
\varphi_{\mathrm{min}}^{2}=\frac{{\Omega^{2}}}{8\lambda_{B}}.
\label{phimin2}
\EE
However we notice that here the mass $\Omega$ must be found
by solution of the coupled gap equations. Thus Eqs.~(\ref{phimin2}) and
(\ref{gap12}) must be regarded as a set of
coupled equations and must be solved together in order to find
the physical values for the correlation lengths and the order
parameter.

Insertion of Eq.~(\ref{phimin2}) into Eq.~(\ref{gap2}) yields a
simple relation for the GL parameter $\kGL$
\BE
\kGL^{2} = \left( \frac{\ell}{\xi}
\right)^{2}=\kappa_0^2 \>
\frac{1}{\displaystyle 1+\frac{I_{0}(\Omega)}{\varphi_{\mathrm{min}}^{2}} }
\label{kappa}
\EE
where $\kappa_0=e_B^2/(8\lambda_B)$ is the mean-field GL parameter
which does not depend on temperature. Equation~(\ref{kappa}) shows that
the GL parameter is predicted to be temperature dependent
through the non trivial dependence of $\Omega$ and $\varphi_{\mathrm{min}}$.
At low temperature, where the order parameter $\varphi_{\mathrm{min}}$ is
large, the deviation from the mean-field value $\kappa_0$ is
negligible. Conversely, close to the critical point, where the
order parameter is vanishing, the correction factor in Eq.~(\ref{kappa})
becomes very important.

\begin{figure}[t]
\centering
\includegraphics[height=\columnwidth,angle=-90]{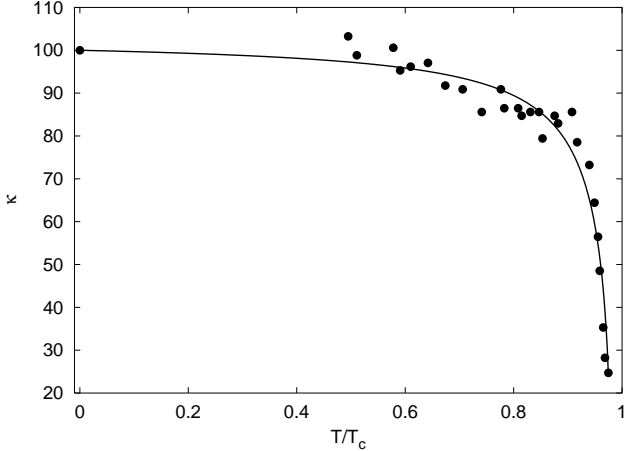}%
\caption{The GL parameter according to Eq.~(\protect\ref{kappa}) for
   $\kappa_0=100$, 
$\xi_0=1.36$ nm, $T_c=121.5$ K and $\Lambda\xi_0=57$ (full line). 
The circles are the experimental data of Ref.~\protect\cite{Brandstatter:94} for 
Tl$_{2}$Ca$_{2}$Ba$_{2}$Cu$_{3}$O$_{10}$.}
\label{fig:Brandstatter}
\end{figure}

\section{Comparison with experimental data}
\label{sec:comparison}

In order to make contact with the phenomenology of the high-$T_c$ cuprate
superconductors we need to fix the bare parameters of the GL
action. The standard derivation \cite{Kleinert:89} of the GL action (\ref{gl})
from a microscopic model gives a direct connection between
microscopic {\it first-principle} quantities and phenomenological
bare parameters. The bare coupling $e_B$ turns out to be related to
the elementary charge of fermions, to the critical temperature $T_c$
and to the zero temperature coherence length $\xi_0$ through \cite{Kleinert:89}
\BE
e_{B}=\frac{2e}{\hbar c}\sqrt{k_{B}T_{c}\xi_{0}} .
\EE
The other parameters may be fixed by knowledge of some zero temperature
phenomenological quantities like coherence length and penetration depth.
The bare mass parameter $m_B^2$ may be regarded as a linear function of
temperature
\BE
m_B^{2}=m_{c}^{2}+\left(1-\frac{T}{T_c}\right) (m_{0}^{2}-m_{c}^{2})
\EE
where $m_0^2$ is the value which is required in order to find 
$\Omega=1/\xi_0$ from the gap equation (\ref{gap1}) at $\varphi=\varphi_{\mathrm{min}}$,
and $m_c^2$ is the value of $m_B^2$ at the transition point.
In mean-field approximation $m_c^2=0$, while here the fluctuations shift
the transition point to a non vanishing $m_B^2$ value.
The bare coupling $\lambda_B$ is regarded as a constant, and is
fixed through Eq.~(\ref{gap2}) by requiring that at zero temperature 
(\emph{i.e.,} for $m_B^2=m_0^2$)
the penetration depth $\ell_0=1/\Delta$.
This way the method provides a one-parameter interpolation scheme
for the superconducting properties, as the cut-off $\Lambda$ still
has to be fixed. The cut-off $\Lambda$ is a characteristic energy scale of the
sample, and may be determined by a direct fit of the experimental data.

Unfortunately there are not too many available experimental data on the behaviour
of superconductors close to the critical point. Even for the
   high-$T_c$ cuprate
superconductors the measurement of both coherence length and penetration
depth up to the pre-critical region is not an easy task.
The GL parameter has been reported 
by Brandstatter \emph{et al.}~\cite{Brandstatter:94} as a function of
   temperature 
for the high $T_c$ material 
Tl$_{2}$Ca$_{2}$Ba$_{2}$Cu$_{3}$O$_{10}$ 
($T_c = 121.5$~K, $\kappa_0=100$, $\xi_0=1.36$ nm).
These data are shown in Fig.~\ref{fig:Brandstatter} together with the
   interpolation curve 
obtained by Eq.~(\ref{kappa}) for $\Lambda\xi_0=57$.
We observe that the experimental GL parameter is almost constant
$\kGL\approx\kappa_0$ up to $T=0.8 T_c$ where it starts
decreasing. Thus for $T/T_c>0.8$ the mean-field approximation
breaks down and the thermal fluctuations become important.
On the other hand Eq.~(\ref{kappa}) fits the experimental data
up to 98\% of the critical temperature. Beyond that point there
are neither experimental data nor reliable theoretical
predictions. As we approach the critical point some universal
behaviour should be expected and the role of thermal fluctuations
becomes too important to be dealt with by the present method \cite{nota:1}.

In perturbative calculations, at one-loop order, the unitarity gauge
   is known \cite{Dolan:74} 
to give rise to wrong predictions around the critical point.
Thus it could be argued that, because of the bad ultraviolet behaviour of 
the gauge propagator, higher order diagrams should be included to cancel
the divergences even in the present one loop calculation. 
However, at variance with perturbative approximations, variational
calculations are known to be less sensitive to higher order corrections.
The GEP provides very sensible results even when the one loop effective
potential fails entirely \cite{Stevenson:85}.
Moreover, while the way of dealing with divergences is crucial at the critical
point, the present calculation deals with a pre-critical region where the
cut-off regulator is finite and plays the role of a physical length scale which
is naturally determined by the structure of the condensed matter system.
Thus divergences are not an issue: our unitarity gauge lagrangian must
be regarded as regularized and is known \cite{Dolan:74} to give the same results
predicted by other gauge choices. 

It must be kept in mind that all variational methods are quite sensitive to
the choice of the trial functional. The Gaussian functional is not by itself
gauge invariant, and a gauge dependent result is expected anyway by this
variational method. The question of determining the best gauge choice has been
addressed by Iba\~nez-Meier \emph{et al.} \cite{IbanezMeier:96} who found the Landau gauge 
$(\partial\cdot A)=0$ to be optimal in four dimensions by 
variational arguments. In three dimensions that is equivalent to the transverse 
gauge we started with in section~\ref{sec:GL}. Unfortunately the Gaussian functional also 
breaks the $U(1)$ symmetry
of the lagrangian, and unphysical massive degrees of freedom show up unless the
unitarity gauge is chosen. Thus our gauge prescription must be regarded as a
compromise which allows to make contact with the real phenomenology. The comparison
with experimental data is encouraging in this respect.

As the GEP provides a nice way to interpolate the experimental
data beyond the mean-field regime, we expect the method to
be reliable for the description of symmetry breaking in $3+1$
dimensions where the scalar electrodynamics may be regarded
as a toy model for the standard electro-weak theory. Since
there is no reason to believe that the real universe is very
close to the transition point, the Gaussian method may
be regarded as a valid tool for describing the effects of
fluctuations beyond mean-field approximation.

\bibliographystyle{mprsty}
\bibliography{Angilella,a,b,c,d,e,f,g,h,i,j,k,l,m,n,o,p,q,r,s,t,u,v,w,x,y,z,zzproceedings,notes}

\begin{thebibliography}{10}

\bibitem{Herbut:96}
I.~F. Herbut and Z.~Te\v{s}anovi\'c, Phys. Rev. Lett. {\bf 76},  4588  (1996).

\bibitem{Tesanovic:99}
Z.~Te\v{s}anovi\'c, Phys. Rev. B {\bf 59},  6449  (1999).

\bibitem{Nguyen:99}
A.~K. Nguyen and A.~Sudb\o, Phys. Rev. B {\bf 60},  15307  (1999).

\bibitem{Mo:02}
S.~Mo, J.~Hove, and A.~Sudb\o, Phys. Rev. B {\bf 65},  104501  (2002).

\bibitem{Larkin:02}
A.~Larkin and A.~A. Varlamov,  in {\em Handbook on Superconductivity:
  Conventional and Unconventional Superconductors}, edited by {K.-H.} Bennemann
  and J.~B. Ketterson (Springer Verlag, Berlin, 2002).

\bibitem{Schneider:00}
T.~Schneider and J.~M. Singer, {\em Phase transition approach to high
  temperature superconductivity. Universal properties of cuprate
  superconductors} (Imperial College Press, London, 2000).

\bibitem{Stevenson:85}
P.~M. Stevenson, Phys. Rev. D {\bf 32},  1389  (1985).

\bibitem{Cuccoli:95}
A.~Cuccoli, R.~Giachetti, V.~Tognetti, R.~Vaia, and P.~Verrucchi, J. Phys.
  Cond. Matt. {\bf 7},  7891  (1995).

\bibitem{IbanezMeier:96}
R.~{Iba\~nez-Meier}, I.~Stancu, and P.~M. Stevenson, Z. Phys. C {\bf 70},  307
  (1996).

\bibitem{Mansfield:86}
P.~Mansfield, Nucl. Phys. B {\bf 267},  575  (1986).

\bibitem{Helfand:66}
E.~Helfand and N.~R. Werthamer, Phys. Rev. {\bf 147},  288  (1966).

\bibitem{Brandstatter:94}
G.~Brandstatter, F.~M. Sauerzopf, H.~W. Weber, F.~Ladenberger, and
  E.~Schwarzmann, Physica C {\bf 235},  1845  (1994).

\bibitem{Stancu:90}
I.~Stancu and P.~M. Stevenson, Phys. Rev. D {\bf 42},  2710  (1990).

\bibitem{Consoli:00}
M.~Consoli and P.~M. Stevenson, Int. J. Mod. Phys. A {\bf 15},  133  (2000).

\bibitem{Kleinert:89}
H.~Kleinert, {\em Gauge Fields in Condensed Matter} (World Scientific,
  Singapore, 1989).

\bibitem{Stevenson:81}
P.~M. Stevenson, Phys. Rev. D {\bf 23},  2916  (1981).

\bibitem{nota:1}
In three dimensions it is well known that the GEP predicts a first order
  transition and fails to predict the details of the transition point.

\bibitem{Dolan:74}
L.~Dolan and R.~Jackiw, Phys. Rev. D {\bf 9},  3320  (1974).

\end{thebibliography}

\end{document}